\documentstyle[prl,aps,epsfig,twocolumn]{revtex}

\newcommand{\be}{\begin{eqnarray}}
\newcommand{\ee}{\end{eqnarray}}
\newcommand{\nn}{\nonumber}

\newlength{\textwidthm}
\begin{document}
\title{A stationary source of non-classical or
entangled atoms}
\author{M.~Fleischhauer$^1$ and S. Gong$^{1,2}$}
\address{$^1$ Fachbereich Physik, Universit\"at Kaiserslautern,
Erwin Schr\"odinger Str., D-67663 Kaiserslautern, Germany\\
$^2$ Laboratory of High-Intensity Optics, Shanghai Institute of Optics
and Fine Mechanics, Shanghai 201800, China}
\date{\today}
\maketitle
\begin{abstract}
A scheme for generating continuous beams of atoms in non-classical
or entangled quantum states is proposed 
and analyzed. For this the recently suggested transfer technique
of quantum states from light fields to collective 
atomic excitation by Stimulated Raman adiabatic passage 
[M.Fleischhauer and M.D. Lukin, Phys.Rev.Lett. {\bf 84}, 5094 (2000)]
is employed and extended to matter waves.
\end{abstract}

\nobreak
Since the experimental realization of Bose-Einstein condensation 
of atoms \cite{BEC} much effort has been put into the generation
of coherent matter waves (atom-lasers) 
\cite{atom_laser}.  Besides being of fundamental interest, 
coherent matter waves are anticipated to lead to
a substantial increase of interferometer sensitivities 
as compared to their optical 
counterparts due to the much smaller wavelength. On the other hand
particle sources have a significantly smaller brightness
than lasers and consequently a much higher level of shot-noise. 
By now experimental techniques have 
advanced to the point that shot-noise limited
operation is achievable in the laboratory \cite{Santarelli99}
and quantum fluctuations represent a true sensitivity limit for
atom interferometer.
Furthermore recent advances in quantum information science 
lead to an increased
interest in sources of entangled massive particles. 
Some time ago 
 Kitagawa and Ueda \cite{Kitagawa93} as well as 
Wineland et al. \cite{Wineland94} proposed to use atom beams
in spin-squeezed states to reduce the noise level in interferometers 
and atomic clocks. 
While optical techniques to generate non-classical light
are by now well developed \cite{squeezed_light}, 
less progress has been made for massive particles.
We here propose a novel scheme for generating 
{\it continuous beams of atoms in  non-classical
quantum states} extending the recently suggested
quantum-state transfer between light and atoms to
matter waves
\cite{Fleischhauer00,Liu01,Phillips01}.

Several proposals for generating atomic beams with non-classical
quantum correlations have been put forward 
and in part experimentally implemented.
Hald et al. \cite{Hald99} reported about spin squeezing
of trapped atoms created by squeezed-light pumping.
The transfer was however accompanied by spontaneous emission 
and only a limited amount of spin squeezing could be achieved. 
A higher degree of noise reduction was obtained by Kuzmich
et al. \cite{Kuzmich00} by continuous QND measurements on an atomic beam.
Recently Pu and Meystre \cite{Pu00} and 
Duan et al. \cite{Duan00} have proposed to make use of 
collisional interactions in a Bose condensate to create squeezing 
or entangled pairs of atoms. Conceptually related is the 
squeezing and entanglement generation by dissociating diatomic
molecules \cite{downconv}. 
Finally the generation of number squeezing in a BEC in an array of weakly 
linked traps was reported by Orzel et al. \cite{Orzel01}. 

Recently we have proposed a technique to transfer the quantum state
of photon wave-packets to collective Raman excitations of atoms in a
loss-free and reversible manner \cite{Fleischhauer00}. First experiments
\cite{Liu01,Phillips01} have 
confirmed important aspects of this proposal.
It is based on the adiabatic rotation of dark-state polaritons, which are
quasi-particles associated with
electromagnetically induced transparency (EIT),
from a light field to a stationary spin excitation. 
When an optically thick sample 
of 3-level atoms in Raman configuration is irradiated by a strong coherent
Stokes field, the absorption of the pump field is 
suppressed. Associated with this
is a substantial reduction of the group velocity of the pump pulse
\cite{group} which corresponds to a temporary storage of its quantum state
in atomic spins. A complete and persistent transfer 
can however only be  achieved by a {\it dynamical} reduction of 
the group velocity due to 
a decrease of the Stokes field intensity in time.
Although an application of this technique to generate entangled or
squeezed samples of atoms has been proposed
in \cite{Lukin00}, it requires an explicit time dependence and is thus 
limited to {\it pulsed} light. For many applications, in
particular sub-shot noise matter-wave interferometry and
continuous teleportation a
{\it stationary} source of non-classical atoms is desired. 
We here show that a complete transfer is possible 
under stationary conditions in a set-up where atoms move through 
a spatially varying Stokes field creating an explicitly time
dependent interaction in their rest frame. 
 In this way a simple
and robust cw source of atoms in non-classical or entangled
states can be build.

\begin{figure}[ht]
\centerline{\epsfig{file=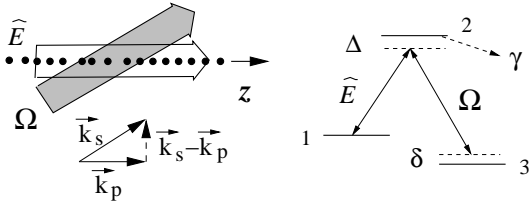,width=7.0cm}}
\vspace*{2ex}
\caption{Beam of 3-level $\Lambda$-type atoms coupled to a classical
field with Rabi-frequency $\Omega(z,t)$ and quantum field $\hat E(z,t)$.
To minimize effect of Doppler-broadening, 
geometry is chosen such that $(\vec k_s-
\vec k_p )\cdot\vec {\rm e}_z\approx 0$.
}
\label{set-up}
\end{figure}

We consider the 1-dimensional model shown in Fig.~\ref{set-up}. A beam of 
$\Lambda$-type atoms with two (meta)-stable lower levels
interacts with a quantized pump and a classical
Stokes field. 
Atoms in different internal states
are described by three bosonic fields $\Psi_\mu(z,t)$ ($\mu =1,2,3$). 
The Stokes field is characterized by the Rabi-frequency
$\Omega(z,t)=\Omega_0(z) \, {\rm e}^{-i\omega_s (t-z/c^\prime)}$ with
$\Omega_0$ taken real, and the 
quantized pump
field by the dimensionless positive frequency component
$\hat E^{(+)}(z,t) ={\cal E}(z,t) 
\, {\rm e}^{-i\omega_p (t-z/c)}$, where
$c^\prime$ denotes the phase velocity projected onto the
$z$ axis. 
The atoms are assumed to enter the interaction region
in state $|1\rangle$.


It is convenient to introduce slowly varying amplitudes,
and a decomposition into velocity classes 
$\Psi_1=\sum_l \Phi_1^l\, {\rm e}^{i(k_lz-\omega_lt)}, 
\Psi_2=\sum_l \Phi_2^l\, {\rm e}^{i[(k_l+k_p)z-(\omega_p+\omega_l) t]}, 
\Psi_3=\sum_l \Phi_3^l\, {\rm e}^{i[(k_l+k_p-k_s)z-
(\omega_p-\omega_s+\omega_l) t]}$, 
where $\hbar k_l$ is the 
momentum of the atoms 
and $\hbar\omega_l=\hbar^2 k_l^2/2m$ the corresponding 
kinetic energy 
in the $l$th velocity class.
$k_p$ and $k_s$ are the wave-vector projection
of pump and Stokes to the $z$ axis.
The atoms shall 
have a narrow velocity distribution around $v_0=\hbar k_0/m$
with $k_0\gg |k_p-k_s|$. All fields are assumed to be in resonance 
for the central velocity class. 
The equations of motion for the matter fields read
\be
\left(\frac{\partial}{\partial t}
+\frac{\hbar k_l}{2 m}\frac{\partial}{\partial z}\right)
\Phi_1^l &=& -i g{\cal E}^+\,\Phi_2^l,\\
\left(\frac{\partial}{\partial t}
+\frac{\hbar k_l}{2 m}\frac{\partial}{\partial z}\right)
\Phi_3^l &=& -i\Omega_0\Phi_2^l - i\delta_l\Phi_3^l,\\
\left(\frac{\partial}{\partial t}
+\frac{\hbar (k_l+k_p)}{2 m}\frac{\partial}{\partial z}\right)
\Phi_2^l &=& -(\gamma+i\Delta_l)\Phi_2^l \\
&& -i\Omega_0\Phi_3^l 
-ig{\cal E}\Phi_1^l +F_2^l,\nonumber
\ee
where $g$ is the atom-field coupling constant and 
$\Delta_l\approx \hbar k_l k_p/m +(\omega_{21}-\omega_p)$
and $\delta_l\approx \hbar k_l (k_p-k_s)/m +(\omega_{31}-
\omega_p+\omega_s)$ are the single and two-photon detunings.
Here second derivatives of the slowly-varying
amplitudes were neglected and sufficiently slow spatial
variations of ${\cal E}$ and $\Omega_0$ assumed.
 $\gamma$ denotes the loss rate out of the
excited state and $F^l_2$ the corresponding Langevin noise operator.
The propagation equation for the elm. field reads
\be
\left(\frac{\partial}{\partial t} +c\frac{\partial}{\partial z}\right)
{\cal E}(z,t)= -ig\Psi_1^{+}(z,t)\Psi_3(z,t).
\ee
The classical Stokes field is taken to be much stronger than the
quantum pump and is assumed undepleted. 
 
In the following we will omitt all Langevin noise terms $F_2^l$ as well
as homogeneous contributions to the solutions. Thus the 
operator relations derived are only valid when taken in normal-ordered
correlation functions.

Consider a stationary input of atoms in state $|1\rangle$, i.e.
$\Psi_1(0,t)=\sqrt{n}$, where $n$ is the constant total density of atoms.
In the limit of a weak quantum field and weak atomic excitation
one finds 
\be
&&\Phi_1^l(z,t)\approx \Phi_1^l(0,t- 2m z/\hbar k_l)
=\sqrt{n}\,\xi_l\, {\rm e}^{-i\varphi_l(z,t)},
\ee
where $\sum_l\xi_l=1$ and $\varphi_l\equiv(k_lz-\omega_lt)$. 
Furthermore
\be
&&\Phi_3^l(z,t) = -\frac{g {\cal E}}{\Omega_0}\sqrt{n}
\,\xi_l\, {\rm e}^{-i\varphi_l(z,t)}
 +\nonumber\\
&&\quad+\frac{i}{\Omega_0(z)}\biggl(\frac{\partial}{\partial t} 
+\frac{\hbar(k_l+k_p)}{2m}\frac{\partial}{\partial z}
+i\Delta_l+\gamma\biggr)\Phi_2^l(z,t),\label{Phi2_gen}\\
&&\Phi_2^l(z,t) =\frac{i}{\Omega_0(z)}\biggl(\frac{\partial}{\partial t} 
+\frac{\hbar k_l}{2m}\frac{\partial}{\partial z}
+i\delta_l\biggr)\Phi_3^l(z,t).
\ee

First the case of perfect two-photon resonance for all
atoms shall be discussed, i.e. $\delta_l\equiv 0$. 
Here one can invoke an adiabatic
approximation, leading to 
\be
\Phi_3^l(z,t) &=&  -\frac{g{\cal E}(z,t)}{\Omega_0(z)} 
\sqrt{n}\,\xi_l\,{\rm e}^{-i\varphi_l(z,t)},
\label{Phi2}\\
\Phi_2^l(z,t) &=&  -i\frac{g \sqrt{n}}{\Omega_0(z)}\xi_l
{\rm e}^{-i\varphi_l(z,t)}
\biggl(\frac{\partial}{\partial t} 
+v_l\frac{\partial}{\partial z}\biggr)
\frac{{\cal E}(z,t)}{\Omega_0(z)},
\ee
with $v_l=\frac{\hbar k_l}{2m}$. 

Substituting the latter result into the equation of motion for the radiation
field yields
\be
&&\left[\left(1+\frac{g^2 n}{\Omega_0^2(z)}\right)
\frac{\partial}{\partial t} +c\left(1+\frac{g^2 n}{\Omega_0^2(z)}
\frac{v_0}{c}\right)\frac{\partial}{\partial z}\right]
\, {\cal E} =\nn\\
&&\qquad \qquad\qquad= \frac{g^2 n}{\Omega_0^2(z)} 
v_0 \left(\frac{\partial}{\partial z}
\ln\Omega_0(z) \right)\, {\cal E},\label{E}
\ee
with $v_0 \equiv \sum_l \xi_l v_l $. 

Thus the quantum pump field propagates with a group velocity
\be
v_{\rm gr}=c\frac{\Bigl(1+\frac{g^2 n}{\Omega_0^2(z)} \frac{v_0}{c}\Bigr)
}{\Bigl(1+\frac{g^2 n}{\Omega_0^2(z)}\Bigr)}
\ee
which approaches $v_0$ if
$\Omega_0 \rightarrow 0$. Note that $v_0>0$ was assumed here. In the case
of atoms moving against the direction of light propagation, i.e. 
for a negative $v_0$, a vanishing and
even a negative value of $v_{\rm gr}$ can arise. 
Moving the medium against the  very small group velocity would
effectively freeze or redirect the light pulse \cite{move}
(note that Galilean laws
apply since $|v_{\rm gr}|\ll c$).
In reality, however,
 non-adiabatic effects and associated losses prevent the latter
to happen \cite{Rochester}.

The r.h.s. of eq.(\ref{E}) describes a reduction / enhancement
due to stimulated Raman adiabatic passage in a spatially varying Stokes
field. 
It can be seen that this is only possible if $v_0\ne 0$
in accordance with the observation in \cite{Fleischhauer00} that a 
unidirectional transfer of excitation requires an explicit time
dependence. For non-vanishing $v_0$ a
space-dependent Stokes field in the laboratory frame 
is equivalent to a time-dependent field in the rest frame of the atoms.

\begin{figure}[ht]
\centerline{\epsfig{file=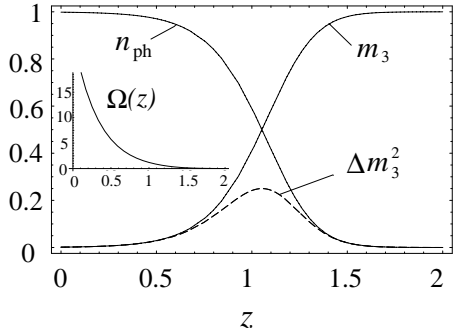,width=6.0cm}}
\vspace*{2ex}
\caption{Average values and fluctuations of 
photons and atoms in state $3$ traversing 
a plane at position $z$ in a given time interval
normalized to input photon number.
Input light is in Fock-state.
$\hat n_{\rm ph}\equiv c\int_T{\rm d}t\, {\cal E}^\dagger(z,t) {\cal E}(z,t)$,
$\hat m_3 \equiv v_0\int_T{\rm d}t\, \Psi_3^\dagger(z,t) \Psi_3(z,t)$.
Insert shows $\Omega(z)$ in units of $g\sqrt{n v_0/c}$.
}
\label{transfer-fig}
\end{figure}

Eq.(\ref{E}) has the simple solution
\be
{\cal E}\Bigl(z,t\Bigr) ={\cal E}\Bigl(0,t-\tau(z)\Bigr)\frac{\cos\theta(z)}
{\cos\theta(0)},
\ee
where $\tau(z)=\int_0^z\!\!{\rm d}z^\prime\, v_{\rm gr}^{-1}(z^\prime)$
and we have introduced the mixing angle $\theta(z)$ according to
$\tan^2\theta(z)\equiv
\frac{g^2 n}{\Omega_0^2(z)} \frac{v_0}{c}$. If $\Omega_0(z)$ is
a sufficiently slowly, monotonically decreasing function 
of $z$ which approaches zero,
the amplitude of the pump field decreases to zero as well. 
At the same time one finds from (\ref{Phi2}) for 
$\Phi_3\equiv \Psi_3\, 
{\rm e}^{-i\{(k_p-k_s)z-(\omega_p-\omega_s)t\}}$:
\be
\Phi_3(z,t)
=-\sqrt{\frac{c}{v_0}}\, \tan\theta(z) {\cal E}\bigl(z,t\bigr)
\ee
If at the input of the interaction region $\theta(0)=0$ and at the
output $\theta(L)= \pi/2$ this yields
\be
\Phi_3(L,t) = -\sqrt{\frac{c}{v_0}}\, {\cal E}\bigl(0,t-\tau),\label{result}
\ee
with $\tau=\int_0^L{\rm d}z v_{\rm gr}^{-1}(z)$.
The factor $\sqrt{c/v_0}$ accounts for the fact that the input
light propagates with velocity $c$ while the output matter field 
propagates only with $v_0$.
As can easily be seen, the input flux of photons is thus equal to the output
flux of atoms in state $|3\rangle$:
$ c \, \langle{\cal E}^+
{\cal E}\rangle_{\rm in} = {v_0} \,
\langle\Psi_{3}^+ \Psi_3\rangle_{\rm out}$.
Eq.(\ref{result}) is the main result of the paper. It shows that 
in the present set-up the quantum properties
of an input electromagnetic field can be completely transferred
to an atomic beam.
This is illustrated in Fig.~\ref{transfer-fig}, where the average value
and fluctuations of the photon number ${\hat n}(z)$ and
the number of atoms in state $3$, ${\hat m}_3(z)$ passing a plane at
position $z$ during a certain time interval are shown. 
The exchange of photons into state-$3$ atoms is apparent. Due to
the incomplete transfer of excitations
atom-number fluctuations reach a maximum for
intermediate $z$, but $\langle\Delta m_3^2\rangle_{\rm out}\, \to\,
\langle\Delta n_{\rm ph}\rangle_{\rm in}$.


Thus continuous matter waves with non-classical quantum correlations
can be generated out of cw non-classical light. 
Since
the mapping technique can be applied to separate Raman transitions
at the same time, it is also possible to transfer entanglement
from a pair of cw light beams as generated for example 
in parametric down-conversion to a pair of atomic beams.

In the derivation of the above result several  
approximations have been invoked. In the following the
validity of those  will be discussed 
in more detail.
The first approximations made are that of a weak quantum pump field
and weak atomic excitation. It can easily be seen from eq.(\ref{Phi2})
that the ratio of the intensity of the pump to the 
Stokes field is given by the ratio of the atomic number density 
in state $|3\rangle$ to the total number density, even when
the Stokes field goes to zero:
$g^2 \langle {\cal E}^+{\cal E}\rangle/\Omega_0^2
=\langle\Psi_3^+\Psi_3\rangle/n$.
It is thus sufficient to fulfill the condition of weak atomic 
excitation, which requires that the input flux of atoms is much
larger than the input flux of pump photons. This condition 
requires some experimental efforts and may not be easy to
satisfy. On the other hand rather high
flux densities of atoms can be achieved in supersonic
beam configurations or (with narrow velocity distributions) in 
atom lasers. 

A second assumption made is that of 
perfect two-photon resonance.
This condition can only be fulfilled if either the velocity
spread of the atoms is extremely small or if the
relative wave-vector  of pump and Stokes fields projected onto
the $z$ axis vanishes. Both conditions are not easy to
satisfy. They require either a 
substantial level of longitudinal cooling, i.e. a coherent
source of input atoms or a careful design of pump and Stokes
field geometry. 
For a quantitative analysis
a non-vanishing but constant value of $\delta_l=\delta$
is considered in the following. 
In this case there is a contribution to $\Phi_2^l$ even in 
lowest order of the adiabatic expansion. 
\be
\Phi_2^l &\rightarrow &  
\Phi_2^l+ \frac{\delta\Omega_0}{\Omega_0^2-\delta(\Delta-i\gamma)}
\frac{g{\cal E}}{\Omega_0} 
\sqrt{n}\,\xi_l\,{\rm e}^{-i(k_lz-\omega_lt)}
\ee
which gives rise to an additional dissipative loss term in the
equation for ${\cal E}$
\be
&&\left[\left(1+\frac{g^2 n}{\Omega_0^2(z)}\right)
\frac{\partial}{\partial t} +c\left(1+\frac{g^2 n}{\Omega_0^2(z)}
\frac{\bar{v}}{c}\right)\frac{\partial}{\partial z}\right]
\, {\cal E} =\qquad\nn\\
&&\qquad= \frac{g^2 n}{\Omega_0^2(z)}
\bar{v} \left(\frac{\partial}{\partial z}
\ln\Omega_0(z) \right)\, {\cal E} -\label{E_loss}\\
&&\qquad\quad - \frac{g^2 n}{\Omega_0^2(z)}\, 
\left(\frac{\delta^2 \Omega_0\gamma}
{(\Omega_0^2-\delta\Delta)^2+\delta^2\gamma^2}\right) 
\, {\cal E} +\cdots\nonumber
\ee
where the dots indicate additional 
imaginary terms that affect only the phase of 
${\cal E}$.

To estimate the influence of the dissipative term
$\Delta=0$ is assumed.
Integrating the field equation yields for the
loss factor $\eta$ of the field amplitude
(${\cal E}\, \to\, \eta\, {\cal E}$)
\be
\eta=\exp\left\{-\alpha \int_0^1\!\!{\rm d}\zeta
\frac{\cos^2\theta(\zeta) x^2}{\cot^4\theta(\zeta) + x^2}\right\},
\label{loss-factor}
\ee
with $\zeta=z/L$. Here $\alpha\equiv g^2 n L/\gamma c$ is the 
opacity of the medium in the
absence of EIT and $x\equiv \delta \gamma/g^2 n \frac{v_0}{c}$ is
a dimensionless quantity characterizing the two-photon detuning. 
The cosine of the mixing angle is monotonically decreasing
from some initial value to zero over the interaction length $L$.
Assuming $x\ll 1$ one can give an upper limit to 
the integral in eq.(\ref{loss-factor}), by replacing the integrand by its
maximum value, which is achieved when $\cos^2\theta\approx |x|\ll 1$. 
This gives the very good estimate
$\eta \ge \exp\{-\alpha |x|/2\}$. Thus in order to neglect the 
influence of dissipative losses, it is sufficient that 
\be
|\delta|\, \frac{L}{v_0}\ll 1.\label{delta-cond}
\ee
%
%
A two-photon detuning can result for example from a residual
Doppler shift of the $1-3$ transition. 
If $\Delta v$ denotes the difference of the velocity in $z$
direction with respect to the resonant
velocity class, the corresponding two-photon detuning reads
$\delta = \Delta v (\vec k_p-\vec k_s)\cdot {\vec e}_z$.
In this case (\ref{delta-cond}) translates into
\be
\frac{|\Delta v|}{v_0}\ll \frac{1}{ (k_p-k_s) L}.\label{v-cond}
\ee
It can be seen from (\ref{v-cond}) that the geometry of the
set-up should be chosen in such a way, that the beat-note wave-vector
has a minimal projection to the $z$ axis. Combining this with
the requirement of a spatially decreasing Stokes field $\Omega_0(z)$ is 
experimentally difficult, but possible in principle. 
Furthermore other schemes of quantum state transfer that use
adiabatic sweeping of the two-photon detuning trough resonance 
rather than an adiabatically varying
Stokes intensity may be employed \cite{Oreg}.

For a monochromatic quantum pump field the conditions
for adiabaticity can easily be obtained from those
in a stationary medium with time dependent Stokes field
\cite{Fl01} by a simple frame transformation. 
This yields
\be
\gamma \int_0^L\!\!{\rm d}z \frac{{v_0} \bigl(\theta^\prime(z)\bigr)^2}{
g^2 n +\Omega_0^2(z)}\ll 1.
\ee
Setting $\theta^\prime(z)\sim 1/L$ this results in
a lower limit for the beam opacity $\alpha =g^2 n L/\gamma c$
in the absence of EIT
\be 
\alpha \gg \frac{v_0}{c}.\label{cond-v}
\ee
Since $v_0\ll c$, a value of $\alpha$ much less than unity will
be sufficient to guarantee total adiabatic transfer.
Note that $v_0$ cannot be arbitrarily small, however, since the
atom flux at the input has to be much larger than the input
photon flux.  

In summary we have shown that a complete and loss-free
transfer of quantum properties from a cw light field
to a continuous beam of atoms is possible using a
recently proposed technique based on Raman adiabatic transfer. 
The combination of a space-dependent Stokes field with a 
finite momentum of the input matter wave leads to a 
time-varying Stokes field in the rest frame of the atoms. 
In this way continuous and monochromatic matter-waves
in non-classical or entangled quantum states can be generated
out of light fields with corresponding properties. 


The financial support of the DFG and the Alexander-von-Humboldt
foundation is gratefully acknowledged. S.G. thanks K. Bergmann
for the hospitality during his stay in Kaiserslautern 
and both authors thank him for stimulating discussions.


\def\etal{\textit{et al.}}

\end{document}